\documentclass[runningheads,orivec]{llncs}
\usepackage[T1]{fontenc}
\usepackage{graphicx}
\usepackage{booktabs}
\usepackage{graphicx}
\usepackage{amsmath,amssymb}
\usepackage{mathtools}
\usepackage{amsfonts,bm}
\newcommand{\halfspace}{\kern 0.1em}
\usepackage{dsfont}
\usepackage{color}
\usepackage{cite}
\usepackage{enumerate}

\usepackage{wrapfig}
\usepackage{capt-of}

\usepackage{hyperref}
\usepackage{url}

\newcommand{\V}{\mathcal{V}}

\newcommand{\E}{\mathbb{E}}

\renewcommand{\S}{\mathcal{S}}
\newcommand{\U}{\mathcal{U}}

\newcommand{\bbf}{\mathbf{b}}
\renewcommand{\P}{\mathcal{P}}
\newcommand{\Tau}{\mathcal{T}}
\newcommand{\zbf}{\mathbf{z}}
\newcommand{\qbf}{\mathbf{q}}
\newcommand{\etabf}{\bm{\eta}}
\newcommand{\ebf}{\mathbf{e}}

\newcommand{\true}{\mathrm{true}}
\newcommand{\one}{\mathbf{1}}
\newcommand{\zero}{\mathbf{0}}

\newcommand{\graph}{\mathcal{G}}
\newcommand{\nodeset}{\mathcal{S}}
\newcommand{\edgeset}{\mathcal{E}}

\newcommand{\defender}{Defender}
\newcommand{\attacker}{Attacker}
\newcommand{\taubf}{\bm{\tau}}
\newcommand{\sigmabf}{\bm{\sigma}}

\newcommand{\paren}[1]{\left({#1}\right)}
\newcommand{\bracket}[1]{\left[{#1}\right]}
\newcommand{\curbracket}[1]{\left\{{#1}\right\}}

\renewcommand{\emptyset}{\varnothing}

\newtheorem{assumption}{Assumption}

\usepackage{tcolorbox}
\newtcolorbox{empheqboxed}{colback=white, 
    colframe=black,
    boxrule=0.25mm,
    width=\columnwidth,
    sharpish corners,
    top=-2mm, 
    left=2pt,
    right=2pt,
    bottom=5pt
}

\NewEnviron{alignsize}[1]{%
\def\scalefactor{#1}
\begin{align}
\BODY
\end{align}
}

\NewEnviron{alignsize*}[1]{%
\def\scalefactor{#1}
\begin{align*}
\BODY
\end{align*}
}

\newcommand{\as}[1]{%
\scalebox{\scalefactor}{$\displaystyle #1$}%
}

\newcommand{\asm}[2]{%
\scalebox{#1}{#2}
}

\usepackage{algorithmic}
\usepackage[ruled,vlined,linesnumbered]{algorithm2e}    
\newcommand{\nonl}{\renewcommand{\nl}{\let\nl\oldnl}}   
\usepackage{listings}

\usepackage{environ}

\NewEnviron{resizealign}{\sbox0{
    $\begin{matrix}\displaystyle\BODY\end{matrix}$}%
  \sbox1{$(\theequation)$}%
  \sbox2{\parbox{\dimexpr \wd0 + 2\wd1}%
    {\begin{align*}\BODY\end{align*}}}
  \noindent\resizebox{\textwidth}{!}{\usebox2}%
}

\allowdisplaybreaks
\begin{document}
\title{Asymmetric-Information Resource Allocation Games: An LP Approach to Purposeful Deception}
\titlerunning{Deceptive Resource Allocation Game}
%
\author{Longxu Pan\inst{1}\thanks{Corresponding author.}
\and
Yue Guan\inst{1}
\and
Daigo Shishika\inst{2}
\and
Panagiotis Tsiotras\inst{1}
}
\authorrunning{L. Pan et al.}
%
\institute{Georgia Institute of Technology, Atlanta GA 30332, USA \and
George Mason University, Fairfax VA 22030, USA
}
\maketitle              
\begin{abstract}
In this work, we introduce the Deceptive Resource Allocation Game (DRAG), which studies purposeful deception within a Bayesian game framework. 
In DRAG, a \defender{} allocates resources across the true asset along with several decoys to influence an \attacker{}’s beliefs and actions, with the intent to divert the \attacker{} away from the true asset. 
We seek to characterize \emph{purposeful deception}, whereby the \defender{} deceives only when doing so improves its performance.
To this end, we solve for the Perfect Bayesian Nash Equilibrium (PBNE) of the corresponding game. 
We show that, despite the coupled belief–policy interdependence, the problem admits an efficient, non-iterative linear programming formulation. 
Numerical results demonstrate that the resulting policies naturally balance performance optimality and belief manipulation, giving rise to purposeful and emergent deceptive behaviors.
\keywords{Multi-Agent System  \and Game Theory \and Decision Making.}
\end{abstract}
%
%
%


\section{Introduction}
Deception plays a fundamental role in domains such~as military operations~\cite{lloyd2003art}, security~\cite{almeshekah2016cyber}, and autonomous systems operating under adversarial influence~\cite{nikitas2022deceitful}.
Existing policies~\cite{chen2024deception} optimize surrogate deception metrics at the expense of performance, leading to ``deception for the sake of deception.''
A common framework that balances deception and performance is asymmetric-information games, where agents possess unequal knowledge about the underlying environment and strategically manipulate observable signals to influence an opponent's belief and decision-making.

A well-studied line of work includes \emph{goal recognition} (GR) and \emph{deceptive path planning} (DPP). 
GR~\cite{Wayllace2016GoalRD, Pereira2017LandmarkBasedHF,ramirez2011goal} focuses on inferring an agent’s intent from observed trajectories, while DPP~\cite{karabag2021deception, rostobaya2025deceptive, rostobaya2023deception} studies how an agent can reach its goal while misleading an observer about its intent. 
Recent DPP works~\cite{rostobaya2025deceptive, rostobaya2026linear} address the limitation of surrogate deception metrics~\cite{chen2024deception} using a Bayesian game formulation.
In these settings, an \attacker{} is privately assigned a goal from a set of candidates, while the \defender{} seeks to identify the true goal by observing the \attacker{}'s trajectory and allocate resources accordingly.
However, existing approaches are limited to settings where the informed agent fully controls the observable signal and, in some cases, to only two candidate goals.
\begin{figure}[t]
    \centering
    \includegraphics[width=0.95\linewidth]{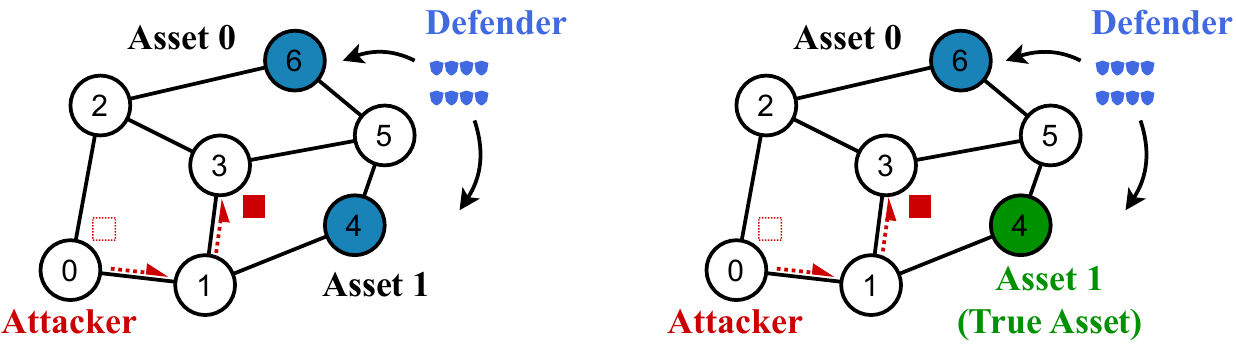}
    \caption{
    An example of DRAG from the perspectives of the \attacker{} (\emph{left}) and \defender{} (\emph{right}).
    }
    \label{fig:DRAG-example}
    \vspace{-0.2in}
\end{figure}

In this paper, we study the Deceptive Resource Allocation Game (DRAG), a natural \emph{dual} of DPP, in which a \defender{}, informed of the true asset, allocates limited resources across candidate assets to mislead the \attacker{} (see Fig.~\ref{fig:DRAG-example}).
Similar in spirit to DPP, the \defender{} in DRAG must balance two competing objectives: allocating resources to the true asset to optimize performance and allocating resources to decoys to manipulate the opponent’s belief. 
This trade-off gives rise to \emph{purposeful} deception, where deception is employed only when it improves the \defender{}'s performance, instead of merely inducing uncertainty or bias to the \attacker{}'s belief.  

In DRAG, the \defender{} influences the \attacker{}'s belief via its allocation actions, while the \attacker{} controls the game state and, in turn, influences the \defender{}'s allocations.
Consequently, the observable signal arises from the interaction among agents, thereby coupling belief and strategic behavior in a way absent from prior DPP formulations.
Thus, unlike DPP~\cite{rostobaya2025deceptive, rostobaya2023deception}, the observable signal in DRAG is \emph{jointly controlled} by both players. 
%
%
To formalize this interaction, we adopt a Bayesian game framework and solve for Perfect Bayesian Nash Equilibria (PBNE), in which agents optimize performance while accounting for belief manipulation. 
Prior works~\cite{lichunLP, guan2025strategic} have proposed linear programming (LP) formulations for computing PBNE when the informed player \emph{fully controls} the observable signal. 
Despite the jointly-controlled signal, we show that DRAG still admits an LP formulation on the game tree.

\noindent
\textit{Contributions:}
The contributions of this work are threefold. 
First, we introduce the DRAG problem, a novel dual formulation of the GR and DPP problems.
Second, we show that the informed player (the \defender{}) can influence the uninformed player's belief (the \attacker{}) by shaping the distribution of observable allocation histories, thereby affecting the \attacker{}'s decisions.
Third, we present an efficient LP formulation on the DRAG game tree, enabling the computation of PBNE.
Our LP approach is general and applies to arbitrary graphs with multiple candidate goals, extending prior DPP work~\cite{rostobaya2025deceptive, rostobaya2023deception}, which is restricted to two-goal settings.
Numerical results demonstrate that the resulting PBNE policies exhibit purposeful deceptive behaviors by balancing deception and performance.

The notation used in this paper is fairly standard.
$\Delta(E)$ denotes the set of probability distributions over a finite set $E$.
For a finite set $\Theta$, $[f(\theta)]_{\theta\in\Theta}$ denotes the vector in $\mathbb{R}^{|\Theta|}$ whose entries are given by $f\!\paren{\theta}$ for each $\theta \!\in \!\Theta$.
Unless otherwise specified, we abbreviate it as $\bracket{f\!\paren{\theta}}_\theta$.
We use $\bracket{0\!:\!T}$ to denote the discrete set $\{0,\!1,\ldots,T\}$.
%
For a random variable $X$ and its realization $x$, the probability of the event $\{X\!=\!x\}$ is denoted as $\P\!\paren{X=x}$.
When the random variable is clear from the context, we use the shorthand $\P\!\paren{x}$.

\section{Problem Formulation}
The Deceptive Resource Allocation Game (DRAG) is a discrete-time dynamic zero-sum game between a \defender{} and an \attacker{} over a graph $\graph = (\nodeset, \edgeset)$,
where $\nodeset$ and $\edgeset$ are the node set and the edge set, respectively.
Let $\Theta \subseteq \nodeset$ denote a set of candidate asset nodes. 
At the start of the game, Nature selects an asset $\theta_\true \!\in\! \Theta$ according to the prior $\bbf_0 \!\in\! \Delta\!\paren{\Theta}$ and privately reveals it to the \defender{}, who aims to protect it, without disclosing to the \attacker{}. 
The \defender{} is of type $\theta$ if the candidate asset $\theta$ is selected by Nature.

At each time step, the \attacker{} moves on the graph $\graph$, while the \defender{} allocates resources to one candidate asset in $\Theta$.
The \defender{} aims to maximize the resources accumulated at $\theta_\true$, while the \attacker{} seeks to minimize the same.
By allocating to decoy assets, the \defender{} can strategically mislead the \attacker{}’s inference of $\theta_\true$, thereby delaying its progress toward the true asset and gaining additional time to reinforce it.
Conversely, the \attacker{} must infer $\theta_\true$ from the observed allocation history and reach it within a prescribed horizon $T$, while minimizing the resources accumulated to $\theta_\true$.
Formally, the DRAG problem is defined as
\[
    \texttt{DRAG} = \langle 
    \nodeset, \edgeset, \Theta, r, m, s_0, \bbf_0,T
    \rangle,
\]
whose components are described in detail below.

\paragraph{State and Action Spaces:}

The game state is given by the \attacker{} position $s_t\! \in\! \nodeset$. 
At time $t$, the \attacker{} selects an edge $u_t \!=\! (s_t, s_{t+1}) \!\in\! \edgeset$ and attempts to move to the next node $s_{t+1}$. 
Thus, the state space is $\nodeset$, and the \attacker{} action space at node $s$ is $\mathcal{U}(s)$, the set of outward edges at $s$.
The \defender{} action $v_t \in \V$ specifies the asset node to which resources are allocated at time $t$. 
Since the \defender{} allocates to only one candidate asset at each time step, its action space $\V$ coincides with $\Theta$.
%
%
The edge weight $w(u_t) > 0$ represents the amount of resources the \defender{} can allocate while the \attacker{} traverses edge $u_t$%
\footnote{
If edge weights represent travel time, then $w(u_t)$ corresponds to the amount of resources allocated by the \defender{} with a unit allocation rate.
}.

\paragraph{Dynamics:}

At the start of the game, Nature selects the true asset $\theta_{\text{true}}$ from the candidate set $\Theta$ according to a commonly known prior $\bbf_0 \!\!\in\!\! \Delta(\Theta)$.
The selected asset is privately revealed to the \defender{}.
With $\theta_{\text{true}}$ fixed throughout the game, the \attacker{} starts from an initial state $s_0 \!\in\! \S$.
All assets are initially unallocated.

At time $t$, both players act \emph{simultaneously}. 
After the \attacker{} takes action $u_t \!\in \U\!\paren{s_t}$ at $s_t$, it transitions to a neighboring node $s_{t+1}$ with probability $\P\!\paren{s_{t+1} \mid s_t, u_t}$.
Concurrently, the \defender{} takes action $v_t\! \in\! \V$, thereby allocating resources to asset $v_t$.
The game terminates when the \attacker{} reaches \emph{any} asset in $\Theta$ or when the horizon~$T$ is reached.
For notational simplicity, we use the shorthand $\P_{\!s_t, u_t}^{s_{t+1}}$ to denote $\P\!\paren{s_{t+1} \vert s_t, u_t}$.

\paragraph{Information Structure:}

All game parameters are common information to both players.
However, the selected asset $\theta_\true$ is only revealed to the \defender{}, leading to an information asymmetry as illustrated in Fig~\ref{fig:DRAG-example}. 
The \attacker{} therefore forms a belief $\bbf_t\! \in\! \Delta(\Theta)$ about the true asset from the observed game history.
At time $t$, both players possess \emph{perfect recall} of the game history $h_t\!=\!\paren{s_0, u_0, v_0, \ldots, s_{t-1}, u_{t-1}, v_{t-1}, s_t}$, containing the \attacker{}'s state-action history and the \defender{}'s action history~\cite{superfair_stochastic_game}. 
%
Let $\mathcal{H}_t$ denote the admissible game history set, and let $|\mathcal{H}_t|$ denote its cardinality.

\paragraph{Behavioral Strategies:}

Since DRAG is a finite extensive-form game with perfect recall, Nash's theorem and Kuhn's theorem guarantee the existence of a Nash equilibrium in behavioral strategies~\cite{Maschler_Solan_Zamir_2013}.
%
Let $\sigma \!\in\! \Sigma$ and $\tau\! \in\! \Tau$ denote the behavioral strategies of the \attacker{} and \defender{}, respectively. 
Because the \defender{} observes its type $\theta$, its behavioral strategy is type-dependent and can be written as $\tau = \curbracket{\tau^\theta}_{\theta\in \Theta}$.
%
At time $t$, the \defender{}'s policy $\tau_t$ maps the history $h_t$ and its type $\theta$ to an action distribution $\taubf_{h_t}^\theta\!\in\!\Delta\!\paren{\V}$
, where $\tau_{h_t}^\theta \!\paren{v_t}$ denotes the probability that \defender{} selects action $v_t$ given its type $\theta$ and the observed history $h_t$.
%
%
Since the \defender{}'s type (i.e., the selected asset) is unobservable to the \attacker{}, the \attacker{}'s policy $\sigma_t$ only takes the history $h_t$ as input and maps to an action distribution $\sigmabf_{h_t}\!\!\in\! \Delta\!\paren{\U\paren{s_t}}$
, where $\sigma_{h_t}\!\paren{u_t}$ represents the probability that \attacker{} selects action $u_t$ given the observed history $h_t$.
%

\paragraph{Reward Structure:}

We consider a general reward function $r(h_t, \theta, u_t, v_t)$ that assigns a running reward at time~$t$ based on the history, the game type, and the actions of both agents. 
The reward is not revealed to the agents during the game and is used only for performance evaluation. 
Therefore, it cannot serve as a signal for inferring the true asset.
For notational simplicity, we introduce the reward matrix $R_{h_t}^\theta \!\!\!\in\!\! \mathbb{R}^{|\U\paren{s_t}|\times |\V|}$, which captures the rewards corresponding to different action pairs.

\subsection{Optimization Objectives}

%
For each type $\theta \!\in\! \Theta$, the \defender{} optimizes its (type-dependent) strategy $\tau^\theta \!=\!\{\taubf_{h_t}^\theta \}_{t=0}^T$ to maximize the following expected accumulated payoff of the type-$\theta$ game,
\begin{equation}
    \scalebox{1}{$V^\theta(h_0, \sigma, \tau^\theta) = \E_{s_0, \theta}^{\sigma, \tau^\theta} \bracket{\sum_{t=0}^{T} r\paren{h_t, \theta, u_t, v_t}}$},
    \label{eqn: type theta value}
\end{equation}
where the superscript indicates that the trajectory distribution is induced by $\paren{\sigma,\tau^\theta}$, while the subscript indicates that the game is initialized at $s_0$ with the type fixed at $\theta$.

Under the zero-sum game assumption and the asymmetric information structure, the \attacker{} optimizes a \emph{single} strategy $\sigma\!=\!\{\sigmabf_{h_t}\}_{t=0}^T$ that minimizes the ex-ante value, 
\begin{equation}
    \scalebox{1}{$V(h_0, \bbf_0, \sigma, \tau) = \E_{s_0, \bbf_0}^{\sigma, \tau} \bracket{\sum_{t=0}^T r\paren{h_t, \theta, u_t, v_t}}.$}
    \label{eqn: ex-ante value}
\end{equation}
Here, $\E_{s_0, \bbf_0}^{\sigma,\tau}$ denotes expectation under the strategy profile $\paren{\sigma, \tau}$, with initial state $s_0$ and $\theta$ drawn according to the prior $\bbf_0$.
We denote the type-$\theta$ and the ex-ante continuation values as $V^\theta\!\paren{h_t, \sigma, \tau^\theta}$ and $V\!\paren{h_t, \bbf_t, \sigma, \tau}$, respectively.
%

\subsection{Perfect Bayesian Nash Equilibrium}

We now formally introduce the \textit{Perfect Bayesian Nash Equilibrium} (PBNE), where the \defender{} maximizes its type-dependent values, and the \attacker{} minimizes the ex-ante value.
The PBNE solution concept consists of two parts: \textit{belief consistency} and \textit{policy sequential rationality} \cite{OuyangCIBPBE, MITOCW-Lec16}.
To formalize belief consistency, we first define the \attacker{}'s belief about the true asset.
\begin{definition}[Belief]
    For $\theta \in \Theta$ and $t \in [0:T]$, the belief $b_t(\theta)$ is the posterior probability that $\theta$ is the true asset given the observed history $h_{t}$, i.e.,
    \vspace{-5pt}
    \begin{equation*}
        b_t\paren{\theta} := \P^\tau\!\paren{\theta\mid h_t},
        \vspace{-8pt}
    \end{equation*}
    where the superscript $\tau$ emphasizes the dependence on the \defender{}'s strategy.
    \label{def:belief-definition}
\end{definition}
At each time $t$, the uninformed \attacker{} updates its belief according to the Bayesian update rule
\vspace{-1em}
\begin{alignsize}{0.95}
    \as{\frac{b_{t-1}(\theta)\,\tau_{h_{t-1}}^{\theta}(v_{t-1})}{\sum_{\theta'} b_{t-1}(\theta')\,\tau_{h_{t-1}}^{\theta'}(v_{t-1})}.}
    \label{eqn:belief-update-rule}
\end{alignsize}
Since the \defender{}'s action $v_{t-1}$ is sampled from the type-dependent policy $\taubf_{h_{t-1}}^\theta$, it serves as the signal for inferring the true asset.
The numerator of~\eqref{eqn:belief-update-rule} is the joint probability of the \defender{} being type $\theta$ and selecting the observed action $v_{t-1}$, while the denominator is the total probability of observing $v_{t-1}$ across all types. 
Thus, \eqref{eqn:belief-update-rule} is the posterior that the \defender{} is of type $\theta$ given $v_{t-1}$.
%

The following assumption and the resulting proposition establish that \eqref{eqn:belief-update-rule} yields the posterior belief $b_t\!\paren{\theta}$ as defined in Definition~\ref{def:belief-definition}.
%
\begin{assumption}[Conditional Independence]
    For $t \in \bracket{1:T}$, conditioned on the observed history $h_t$, the belief over the true asset is independent of the \attacker{}'s immediate action-state pair $\paren{u_{t-1}, s_t}$, i.e.,
    \begin{equation*}
        b_t\!\paren{\theta}:= \P^\tau\!\big(\theta \mid \underbrace{h_{t-1}, u_{t-1}, v_{t-1}, s_t}_{h_t}\big) = \P^\tau\!\paren{\theta \mid h_{t-1}, v_{t-1}},
    \end{equation*}
    for all $\paren{u_{t-1}, s_t} \in \U\!\paren{s_{t-1}} \times \S$.
    \label{assump:belief-cond-indep}
\end{assumption}
%

%

\begin{proposition}
    Under Assumption~\ref{assump:belief-cond-indep}, the Bayesian update rule \eqref{eqn:belief-update-rule} coincides with $b_t\paren{\theta}$, i.e.,
    \begin{equation*}
        b_t\!\paren{\theta} = \frac{b_{t-1}(\theta)\,\tau_{h_{t-1}}^{\theta}(v_{t-1})}{\sum_{\theta'} b_{t-1}(\theta')\,\tau_{h_{t-1}}^{\theta'}(v_{t-1})}.
    \end{equation*}
    \vspace{-10pt}
    \label{prop:belief-update-satisfies-belief-definition}
\end{proposition}
\begin{proof}
    The result follows from the interpretation of \eqref{eqn:belief-update-rule} as $\P^\tau\!\paren{\theta \mid h_{t-1}, v_{t-1}}$, which equals to $b_t\!\paren{\theta}$ under Assumption~\ref{assump:belief-cond-indep}.
    See Appendix~\ref{appendix:belief-update-justification} for details.
\end{proof}
%
%
The PBNE is formally defined as follows.
\begin{definition}
    A strategy profile $(\sigma^*, \tau^*)$ constitutes a PBNE if the following holds for all $t\in\bracket{0:T}$ and all histories $h_t \in \mathcal{H}_t$ with non-zero probability,
    \begin{enumerate}
        \item \textit{Sequential rationality}:
        \vspace{-0.05in}
        \begin{alignsize*}{1}
            \as{\hspace{-0.1in} V(h_t, \bbf_t, \sigma^*\!\!, \tau^*)} ~&\as{\leq V(h_t, \bbf_t, \sigma, \tau^*)}, ~ &&\as{\forall\;\sigma \in \Sigma}, \\
            \as{\hspace{-0.1in} V^\theta(h_t,  \sigma^*\!\!, \tau^{\theta *})} ~&\as{\geq V^\theta(h_t, \sigma^*\!\!, \tau^\theta)}, ~ && \as{\forall\, \theta \!\in\! \Theta, \tau^\theta \!\in\! \Tau^\theta}.
        \end{alignsize*}
        \vspace{-0.2in}
        \item \textit{Belief consistency}: the belief $\bbf_{t}$ is propagated according to the Bayesian update rule in~\eqref{eqn:belief-update-rule}.
    \end{enumerate}
\end{definition}
For brevity, we use $V^*\!\paren{h_t, \bbf_t}$ to denote $V\!\paren{h_t, \bbf_t, \sigma^*, \tau^{\theta*}}$.
Similarly, $V^{\theta*}\!\paren{h_t}$ denotes $V^\theta\!\paren{h_t, \sigma^*, \tau^{\theta*}}$.



\section{Dynamic Programming for Value Functions}
A key challenge in DRAG is the coupling between belief updates and strategic behaviors.
To address this, we first establish a dynamic programming (DP) representation of the ex-ante value $V\!\paren{h_t, \bbf_t, \sigma, \tau}$.
Building on this structure, we show that the equilibrium game value $V^*\!\paren{h_t, \bbf_t}$ admits a min-max DP characterization.
These results enable a sequence of duality-based reformulations on $V^*\!\paren{h_t, \bbf_t}$, through which we construct an LP over the game tree~\cite{koller1996efficient}.

Given the history $h_t \!\in\! \mathcal{H}_t$ and belief $\bbf_t$ at time $t$, the strategy profile $\paren{\sigma, \tau}$ and the transition dynamics induce a probability distribution over future game trajectories.
Under this induced distribution, the type-$\theta$ continuation value $V^\theta \!\paren{h_t, \sigma, \tau^\theta}$ and the corresponding ex-ante value $V\!\paren{h_t, \bbf_t, \sigma,\tau}$ admit the following DP representations

\vspace{-0.2in}
\begin{alignsize}{1}
& \as{V^\theta \!\paren{h_t, \sigma, \tau^\theta} = \sigmabf_{h_t}^\top R_{h_t}^\theta \taubf_{h_t}^\theta + \!\!\!\!\!\sum_{u_t, v_t, s_{t+1}} \!\!\!\!\!\!\sigma_{h_t}\!\paren{u_t} \P^{s_{t+1}}_{s_t, u_t}\,\tau_{h_t}^\theta\! \paren{v_t} V^\theta \!\!\paren{h_{t+1}, \sigma, \tau^\theta}},
\label{eqn:V^theta-DP}
\\
& \as{V \!\paren{h_t, \bbf_t, \sigma, \tau} = \sum_\theta b_t(\theta) \,\sigmabf_{h_t}^\top R_{h_t}^\theta \taubf_{h_t}^\theta}
\label{eqn:V-almost-DP}
\\[-3pt]
& \qquad \qquad \qquad \qquad\as{+ \!\!\!\!\!\sum_{u_t, v_t, s_{t+1}} \!\!\!\!\!\!\sigma_{h_t}\!(u_t)  \underbrace{\sum_\theta \P^{s_{t+1}}_{\!\!s_t, u_t} \bracket{b_t(\theta) \tau_{h_t}^\theta\!\paren{v_t} V^\theta\!\paren{h_{t+1}, \sigma, \tau^\theta}}}_{A}}.
\nonumber
\end{alignsize}
In particular, the continuation term $V^\theta\!\paren{h_{t+1}, \sigma, \tau^\theta}$ in \eqref{eqn:V^theta-DP} and \eqref{eqn:V-almost-DP} denotes the expected accumulated payoff from stage $t+1$ to horizon $T$ under type $\theta$.
%

The terminal values are given by 
\vspace{-0.05in}
\begin{alignsize}{1}
    \as{V^\theta (h_T, \sigma, \tau^\theta)} & \as{= \sigmabf_{h_T}^\top R_{h_T}^\theta \taubf_{h_T}^\theta},
    \label{eqn:V^theta-terminal}
    \\
    \as{V(h_T, \bbf_T, \sigma, \tau)} & \as{= {\textstyle \sum_\theta} \; b_T(\theta) \sigmabf_{h_T}^\top R_{h_T}^\theta \taubf_{h_T}^\theta}.
    \label{eqn:V-terminal}
\end{alignsize}
Note that \eqref{eqn:V-almost-DP} is not a DP in terms of the ex-ante value $V$ alone due to the presence of $V^\theta$ in $A$.  
To resolve this, we leverage the following properties of ex-ante values from \eqref{eqn:V^theta-DP}-\eqref{eqn:V-terminal}.

%
\vspace{-5pt}
\begin{remark}
The ex-ante value is the expectation of the type-$\theta$ values, that is, 
\vspace{-5pt}
\begin{equation}
    \textstyle V\!\paren{h_t, \bbf_t, \sigma, \tau}\!=\!\sum_{\theta} b_t\paren{\theta} V^\theta \!\paren{h_t, \sigma, \tau^\theta}.
    \label{eqn: V = sum_theta b_theta V^theta}
\end{equation}
\label{rmk: V = sum_theta b(theta) V^theta}
\end{remark}
\vspace{-25pt}
%
\begin{remark}
    The ex-ante value is positively homogeneous in the belief, i.e., for $\alpha \geq 0$,
\vspace{-5pt}
    \begin{equation}
        V\!\paren{h_t, \alpha \bbf_t, \sigma, \tau} = \alpha V\!\paren{h_t, \bbf_t, \sigma, \tau}.
        \label{eqn: V(ab) = a V(b)}
    \end{equation}
    \label{rmk: V(a b_t) = a V(b_t)}
\end{remark}
\vspace{-16pt}
%
Rearranging the coefficients of the term $A$ in~\eqref{eqn:V-almost-DP}, we have
\vspace{-5pt}
\begin{alignsize*}{1}
    \as{A~} & \as{= {\textstyle \sum_\theta} \; \P^{s_{t+1}}_{\!\!s_t, u_t} \xi \Big[\scalebox{0.8}{$\displaystyle \frac{b_t(\theta) \tau_{h_t}^\theta\!\paren{v_t}}{\xi}$}\, V^\theta\!\!\paren{h_{t+1}, \sigma, \tau^\theta}}\Big] \\
    & \as{=
    {\textstyle \sum_\theta} \; \P^{s_{t+1}}_{\!\!s_t, u_t} \xi \bracket{b_{t+1}(\theta) V^\theta\!\!\paren{h_{t+1}, \sigma, \tau^\theta}}},
\end{alignsize*}
where $\xi \!=\! \sum_{\theta'} \!b_{t}(\theta')\,\tau_{h_{t}}^{\theta'}(v_{t})$ is the denominator in~\eqref{eqn:belief-update-rule}.
We use Remarks~\ref{rmk: V = sum_theta b(theta) V^theta}-\ref{rmk: V(a b_t) = a V(b_t)} to relate the type-$\theta$ value in term $A$ with the ex-ante value, yielding the following DP in terms of the ex-ante value $V$ alone:
\begin{alignsize}{0.95}
    & \as{V\!\paren{h_t, \bbf_t, \sigma, \tau} = \sum_\theta b_t(\theta)\, \sigmabf_{h_t}^\top R_{h_t}^\theta \taubf_{h_t}^\theta + \!\!\!\!\!\sum_{u_t, v_t, s_{t+1}} \!\!\! \!\!\Big[\sigma_{h_t}\!\paren{u_t} V\big(h_{t+1},\,
    {\alpha\,\bbf_{t+1}},\, \sigma,\, \tau\big)\Big]},  \label{eqn: V DP}
\end{alignsize}
where $\alpha \!= \!\P_{\!s_t, u_t}^{s_{t+1}}\, \xi$.
Since \eqref{eqn: V DP} is derived under the Bayesian belief update rule, the belief in the recursion is consistent with the underlying $\paren{\sigma, \tau}$.
Thus, the PBNE characterization reduces to enforcing sequential rationality on \eqref{eqn: V DP}, yielding the min-max DP characterization in~\eqref{eqn:minmax-V-DP},
\begin{subequations}
    \label{eqn:minmax-V-DP}
    \begin{alignsize}{0.77}
        \as{V^*\!\paren{h_{t}, \bbf_{t}}}&\as{=\min_{\sigmabf_{h_{t}}} \max_{\taubf_{h_{t}}} \bigg\{\sum_\theta b_{t}(\theta)\,\sigmabf_{h_{t}}^\top\,  R_{h_{t}}^\theta \,\taubf_{h_{t}}^\theta + \!\!\!\!\!\!\!\sum_{u_{t}, v_{t}, s_{t+1}}\!\!\!\!\!\! \sigma_{h_{t}}(u_{t}) \, {V^*\!\Big(h_{t+1}, \underbrace{\bracket{\P_{s_{t}, u_{t}}^{s_{t+1}}\, b_{t}(\theta')\, \tau_{h_t}^{\theta'}(v_t)}_{\theta'}}_{\alpha \bbf_{t+1}}\Big)}\bigg\}}
        \label{eqn:minmax-V-DP original}
        \\[-5pt]
        & \as{= \min_{\sigmabf_{h_{t}}} \bigg\{ \!\underbrace{\max_{\taubf_{h_{t}}} \,\sigmabf_{h_t}^\top \Big\{\sum_\theta b_{t}(\theta)\,  R_{h_{t}}^\theta \,\taubf_{h_{t}}^\theta + \Big[\sum_{ v_{t}, s_{t+1}}\!\!\! V^*\!\big(h_{t+1}, \big[\P_{s_{t}, u_{t}}^{s_{t+1}}\, b_{t}(\theta')\, \tau_{h_t}^{\theta'}(v_t)\big]_{\theta'}\big)\Big]_{u_t}\!\Big\}}_{B\paren{\sigmabf_{h_t}}}\!\bigg\},}
        \label{eqn:minmax-V-DP-max-isolated}
    \end{alignsize}
\end{subequations}
where the scaled belief $\alpha \bbf_{t+1}$ is written explicitly in~\eqref{eqn:minmax-V-DP original} for clarity.
The max-min DP follows by symmetry.

\section{\defender{}'s Linear Program}
We now construct the LP that computes the \defender{}'s equilibrium strategy $\tau^*$.
The key idea is to introduce variables that encode the joint effect of belief and \defender{} policies~\cite{lichunLP}, thereby eliminating the nonlinear coupling between belief updates and policy choices.

\subsection{Single-stage LP}

We begin with the single-stage case.
Given $(h_T, \bbf_T)$, the DRAG reduces to a static Bayesian game whose value $V^*\!\paren{h_T, \bbf_T}$ is given by the following LP:
\begin{subequations}\label{LP:defender-single-stage} 
\begin{alignsize}{1}
    \as{V^*\!\paren{h_T, \bbf_T}\,\,} & \as{=  \!\!\!\max_{\asm{0.8}{$\taubf_{h_T},\, \ell_{h_T}$}} \ell_{h_T}}\nonumber\\
    \as{\text{s.t.}\quad}
    & \as{\one^\top \, \taubf_{h_T}^\theta = 1, ~~ \taubf_{h_T}^\theta \geq \zero, \quad\quad\qquad\forall \, \theta \in \Theta,}\\
    & \as{\ell_{h_T} \one\leq {\sum_{\asm{0.6}{$\theta$}}} R_{h_T}^\theta b_T\!\paren{\theta}\taubf_{h_T}^\theta,} \label{constr:single-stage-defender-optimality} 
\end{alignsize}
\end{subequations}
where the scalar $\ell_{h_T}$ represents the worst-case ex-ante value that the \defender{} can achieve under belief $\bbf_T$.
Note that if $\bbf_T$ is scaled by a nonnegative scalar $\alpha$, the LP remains valid, with $b_T\!\paren{\theta}$ replaced by the scaled belief $\alpha \, b_T\!\paren{\theta}$ in \eqref{constr:single-stage-defender-optimality}.

\subsection{Two-stage LP}
We now consider the two-stage DRAG.
Notice in~\eqref{eqn:minmax-V-DP-max-isolated} that the maximization over the \defender{}'s policy can be isolated as an inner problem for a fixed \attacker{} policy.
The term $B\!\paren{\sigmabf_{h_{T-1}}}$ can be reformulated as an LP, dualized, and then embedded back into the overall optimization.

Under~\eqref{eqn:minmax-V-DP}, the equilibrium value $V^*\!\paren{h_{T\!-\!1}, \bbf_{T\!-\!1}}$ is given by a min-max problem that involves the continuation value $V^*\!\paren{h_T, \alpha\bbf_T}$ at time $T$. 
By the single-stage result \eqref{LP:defender-single-stage}, $V^*\!\paren{h_T, \alpha \bbf_T}$ admits the following optimization:
\begin{alignsize}{1}
    \as{V^*\!\paren{h_T, \alpha \bbf_T}\,\,} &  \as{=\! \max_{\asm{0.8}{$\taubf_{h_T},\ell_{h_T}$}} \!  \ell_{h_T}}
    \label{LP:two-stage-V(h_T,.)}
    \\
\text{s.t.} ~~
& \as{\displaystyle \one^\top \taubf_{h_T}^\theta \!=\! 1, \quad \taubf_{h_T}^\theta \geq \zero, \quad\quad\qquad\qquad\qquad \forall \, \theta \!\in\! \Theta},\nonumber\\
&\as{\ell_{h_T} \one\leq{\displaystyle \sum_{\asm{0.6}{$\theta$}}} R_{h_T}^\theta \underbrace{\P_{\!\!s_{T\!-\!1}\!, u_{T\!-\!1}}^{s_T} b_{T\!-\!1}\!\paren{\theta} \tau_{h_{T\!-\!1}}^\theta\!\!\paren{v_{T\!-\!1}}}_{\alpha \, b_T\paren{\theta}} \taubf_{h_T}^\theta}. \nonumber
\end{alignsize}
However, \eqref{LP:two-stage-V(h_T,.)} cannot be solved independently from the optimization at $T\!-\!1$, since $\alpha$ contains $\tau_{h_{T\!-\!1}}^\theta\!\!\paren{v_{T\!-\!1}}$ which is a decision variable at $T\!-\!1$.
Given a fixed $\sigmabf_{h_{T-1}}$,
the term $B\!\paren{\sigmabf_{h_{T-1}}}$ in \eqref{eqn:minmax-V-DP-max-isolated} can be reformulated as the following optimization problem by incorporating \eqref{LP:two-stage-V(h_T,.)} 
\vspace{-3pt}
\begin{subequations}
    \label{LP:two-stage-inner-max-no-z}
    \begin{alignsize}{1}
    \as{B\paren{\sigmabf_{h_{T\!-\!1}}}\!=\hspace{-0.1in}} & \as{\max_{\asm{0.8}{$\substack{\taubf_{h_{T\!-\!1}}
    \\ \taubf_{h_T},\, \ell_{h_T},\, \forall \, h_{T}}$}}  \!\!\!\!\!\sigmabf_{h_{T\!-\!1}}^\top \bigg\{\sum_{\asm{0.6}{$\theta$}} R_{h_{T\!-\!1}}^\theta b_{T\!-\!1}\!\paren{\theta} \taubf_{h_{T\!-\!1}}^\theta + \bigg[\sum_{\asm{0.6}{$v{_{T\!-\!1}, s_{T}}$}} \!\ell_{h_T}\bigg]_{\asm{0.6}{$u_{T\!-\!1}$}}\bigg\}} \nonumber\\[+3pt]
        \as{\mathrm{s.t.} } & \as{\quad\one^\top \taubf_{h_{T\!-\!1}}^\theta \!\!= \!1,} \as{\quad \taubf_{h_{T\!-\!1}}^\theta\! \geq\! \zero}, \as{\qquad\qquad\qquad\qquad\qquad~~~ \forall \, \theta \!\in\! \Theta}, 
        \label{constr:two-stage-inner-max-tau_h_T-probability}\\[+3pt]
        & \forall \, h_{T} \in \mathcal{H}_T, \nonumber\\
        & \as{\quad\one^\top \taubf_{h_T}^\theta \!=\! 1, \quad \taubf_{h_T}^\theta \!\geq \!\zero}, \as{\qquad\qquad\qquad\qquad\qquad\qquad \forall \,\theta \!\in \!\Theta,}
        \label{constr:two-stage-inner-max-tau_h_T-1-probability}\\
        & \as{\quad\ell_{h_T} \one \leq  \sum_{\asm{0.6}{$\theta$}} R_{h_T}^\theta \P_{\!\!\!s_{T\!-\!1}, u_{T\!-\!1}}^{s_T} b_{T\!-\!1}\!\paren{\theta} \tau_{h_{T-1}}^\theta\!\!\paren{v_{T\!-\!1}} \taubf_{h_T}^\theta}.
        \label{constr:two-stage-inner-max-l_h_T-optimality}
    \end{alignsize}
\end{subequations}
Notice that the constraint \eqref{constr:two-stage-inner-max-l_h_T-optimality} involves products between $\tau_{h_{T\!-\!1}}^\theta\!\!\paren{v_{T\!-\!1}}$ and $\taubf_{h_T}^\theta$, which results in a bilinear constraint. 
To restore linearity in~\eqref{LP:two-stage-inner-max-no-z}, we introduce auxiliary variables $\zbf$ that encode the joint effect of belief and policies across stages.
Specifically, for all $\theta \in \Theta$ and histories $h_T \in \mathcal{H}_T$, we define
\begin{equation}
    \label{eqn:two-stage-defender-change-of-var}
    \asm{1}{$\zbf_{h_{T\!-\!1}}^\theta \!\!\coloneqq b_{T\!-\!1}(\theta)\, \taubf_{\!h_{T\!-\!1}}^\theta$}, 
    \quad 
    \asm{1}{$\zbf_{h_T}^\theta \!\coloneqq 
    \scalebox{1}{$\P_{\!\!s_{T\!-\!1}, u_{T\!-\!1}}^{s_T}$} z_{h_{T\!-\!1}}^\theta\!\!\paren{v_{T\!-\!1}} \taubf_{h_T}^\theta$}.
\end{equation}

The variable $z_{h_{T\!-\!1}}^\theta\!\!\paren{v_{T\!-\!1}}$ represents the joint probability of the \defender{} being type-$\theta$ and selecting action $v_{T\!-\!1}$ at history $h_{T\!-\!1}$.
For each successor history $h_T \!= \!(h_{T\!-\!1}, u_{T\!-\!1},$ $v_{T\!-\!1}, s_T)$, the variable $z_{h_T}^\theta\!\paren{v_T}$ represents the joint probability of the \defender{} being type-$\theta$ and transitioning 
$h_{T-1}$
to $h_T$ and then select action $v_T$.
Finally, the policy can be recovered as 
\begin{equation}
\tau^\theta_{h_t}(v_t) = \frac{z^\theta_{h_t}(v_t)}{\sum_{v'} z^\theta_{h_t}(v')}.
\end{equation}

Using the auxiliary $\zbf$ variables, \eqref{LP:two-stage-inner-max-no-z} can be transformed to the equivalent LP under fixed \attacker{} policy $\sigmabf_{h_{T-1}}$,
\vspace{-5pt}
\begin{subequations}
\label{LP:two-stage-inner-max}
\begin{alignsize}{1}
    \as{B\!\paren{\sigmabf_{h_{T-1}}}~}&\as{=\hspace{-0.1in} \max_{\asm{0.8}{$\substack{\zbf_{h_{T\!-\!1}}\\ \zbf_{h_T},\, \ell_{h_T}, \forall\, h_T}$}} \hspace{-0.1in}
    \sigmabf_{h_{T\!-\!1}}^\top 
    \bigg\{\!\sum_{\asm{0.6}{$\theta$}} R_{h_{T\!-\!1}}^\theta \zbf_{h_{T\!-\!1}}^\theta + \bigg[\sum_{\asm{0.6}{$v_{T\!-\!1}\!, s_T$}}\! \ell_{h_T}\bigg]_{\asm{0.6}{$\!u_{T\!-\!1}$}}\!\bigg\}}
    \label{obj:two-stage-inner-max}
    \\
     \as{\text{s.t.} ~~}
     & \as{\quad \one^\top \zbf_{h_{T\!-\!1}}^\theta \!\!= b_{T\!-\!1}\!\paren{\theta},}\quad \as{\zbf_{h_{T\!-\!1}}^\theta \geq \zero}, \as{\quad\quad\quad~~~\forall \, \theta \in \Theta},
     \label{constr:two-stage-inner-max-z_h_T-1-probability}\\[+5pt]
    & \forall \, h_{T} \in \mathcal{H}_T, \nonumber\\ 
    &\as{\quad\one^\top \zbf_{h_T}^\theta \!\!=\! 
    \P_{\!\!s_{T\!-\!1}\!, u_{T\!-\!1}}^{s_T} z_{h_{T\!-\!1}}^\theta\!\!\paren{v_{T\!-\!1}},
    \,\as{\zbf_{h_T}^\theta \geq \zero}, ~~ \forall \, \theta \in \Theta}, \label{constr:two-stage-inner-max-z_h_T-probability}\\
    &\as{\quad\ell_{h_T} \one\leq {\displaystyle \sum_{\asm{0.6}{$\theta$}}} R_{h_T}^\theta \zbf_{h_T}^\theta}. \label{constr:two-stage-inner-max-l_h_T-optimality-z}
    \\[-25pt]\nonumber
\end{alignsize}
\end{subequations}

\noindent
Note that~\eqref{LP:two-stage-inner-max} yields $B\!\paren{\sigmabf_{h_{T-1}}}$ in~\eqref{eqn:minmax-V-DP-max-isolated} for a fixed $\sigmabf_{h_{T-1}}$.
%
Since a feasible LP with its objective bounded in the direction of optimization admits a finite optimal solution and satisfies strong duality~\cite{bertsimas-LPbook}, it suffices to show that~\eqref{LP:two-stage-inner-max} under fixed $\sigmabf_{h_{T\!-\!1}}$ is feasible and upper bounded.
Feasibility follows because \eqref{constr:two-stage-inner-max-z_h_T-1-probability} and~\eqref{constr:two-stage-inner-max-z_h_T-probability} construct non-empty scaled simplices, while $\ell_{h_T}$ can be chosen sufficiently small to satisfy~\eqref{constr:two-stage-inner-max-l_h_T-optimality-z}.
%
One can easily verify the upper-boundedness of its objective~\eqref{obj:two-stage-inner-max} (see Appendix~\ref{appendix:lemma-1-proof}).
%
%
%
%
Dualizing \eqref{LP:two-stage-inner-max} and combining with the outer minimization over $\sigmabf_{h_{T-1}}$ in \eqref{eqn:minmax-V-DP} leads to the following LP for $V^*(h_{T\!-\!1}, \bbf_{T\!-\!1})$
\begin{subequations}
\label{LP:two-stage-overall-min}
\begin{alignsize}{1}
    \as{V^*\!\paren{h_{T\!-\!1},\bbf_{T\!-\!1}}~} & \as{=\hspace{-0.1in} \min_{\asm{0.8}{$\substack{\sigmabf_{\!h_{T\!-\!1}}, \qbf_{h_{T\!-\!1}}\\ \etabf_{h_{T}}, \qbf_{h_T}, \forall \, h_T}$}}  \sum_{\asm{0.8}{$\theta$}} b_{T\!-\!1}\!\paren{\theta} q_{h_{T\!-\!1}}\!\!\paren{\theta}}
    \label{obj:two-stage-overall-min}\\
    \as{\text{s.t.}  ~~}
    &   \as{\forall \, \theta \!\in\! \Theta,\,\forall \,v_{T\!-\!1} \!\in\! \V,}\nonumber\\
    &  \as{\quad \one^\top \sigmabf_{\!h_{T\!-\!1}} = 1, \quad \sigmabf_{\!h_{T\!-\!1}} \geq \zero}, 
    \label{constr:two-stage-overall-min-sigma_h_T-1-probability}\\
    & \as{\quad 
    q_{h_{T\!-\!1}}(\theta) \geq
     \ebf_{v_{T\!-\!1}}^\top R_{h_{T\!-\!1}}^{\theta^\top} \sigmabf_{\!h_{T\!-\!1}}+\!\!\sum_{\asm{0.6}{$u_{T\!-\!1}, s_T$}}\!\! \P_{{\!\!\!s_{T\!-\!1}\!, u_{T\!-\!1}}}^{s_T} q_{h_T}\!\paren{\theta},}
    \label{constr:two-stage-overall-min-q_h_T-1-optimality}
    \\
    &  \as{\forall \, h_T\! \in\! \mathcal{H}_T,\, \forall \, \theta \!\in\! \Theta,\, \forall \, v_T \!\in\! \V,} \nonumber 
    \\
    &\as{\quad \one^\top \etabf_{h_T}\!= \sigma_{h_{T\!-\!1}}\!\!\paren{u_{T\!-\!1}}, \quad \etabf_{h_{T}} \geq \zero}, 
    \label{constr:two-stage-overall-min-eta_h_T-probability}\\
    & \as{\quad  q_{h_T}(\theta) \geq \ebf_{v_T}^\top R_{h_T}^{\theta^\top} \etabf_{h_T},}
    \label{constr:two-stage-overall-min-q_h_T-optimality}
\end{alignsize}   
\end{subequations}
where $\qbf_{h_{T\!-\!1}}, \qbf_{h_T} \!\in\! \mathbb{R}^{|\Theta|}$, and $\etabf_{h_{T}} \!\in\! \mathbb{R}^{|\U\paren{s_T}|}$ are the dual variables, and $\ebf_{v_{T\!-\!1}},\ebf_{v_T}\!\in\!\mathbb{R}^{|\V|}$ are unit vectors with $1$ on the entries corresponding to $v_{T\!-\!1}$ and $v_T$.
As shown later in the \attacker{}’s LP, the vectors $\qbf$ correspond to the \attacker{}’s performance under each \defender{} allocation action, while $\etabf_{h_T}$ characterizes the \attacker{}’s policy at history $h_T$.
%
By strong duality, the dual of~\eqref{LP:two-stage-overall-min} is feasible and admits the same finite value $B\!\paren{\sigmabf_{h_{T\!-\!1}}}$ under fixed $\sigmabf_{h_{T\!-\!1}}$.
Since the additional \attacker{} policy simplex~\eqref{constr:two-stage-overall-min-sigma_h_T-1-probability} is non-empty and the finite reward function lower-bounds the objective~\eqref{obj:two-stage-overall-min}, strong duality extends to the combined minimization LP~\eqref{LP:two-stage-overall-min}.

Dualizing~\eqref{LP:two-stage-overall-min} yields the two-stage \defender{}'s maximization LP

\begin{subequations}
    \label{LP:two-stage-Defender-LP}
    \vspace{-15pt}
    \begin{alignsize}{1}
    \as{V^*\!\paren{h_{T\!-\!1},\,\bbf_{T\!-\!1}}}~ & \as{= \hspace{-0.1in} \max_{\asm{0.8}{$\substack{\zbf_{h_{T\!-\!1}}\!,\, \ell_{h_{T\!-\!1}}\\ \zbf_{h_T}\!,\, \ell_{h_T}\!, \forall \, h_T}$}}  \ell_{h_{T\!-\!1}}}
    \label{obj: two stage Defender}
    \\
    \as{\text{s.t.}~~} 
    & \as{\quad \one^\top \zbf_{h_{T\!-\!1}}^\theta \!\!= b_{T\!-\!1}\!\paren{\theta}}, \as{\quad \zbf_{h_{T\!-\!1}}^\theta \!\!\geq \zero,} \as{\quad\quad\qquad\qquad\forall \, \theta \!\in\!\Theta,}
    \label{constr: two stage defender z_h_T-1 probability}
    \\
    & \as{\quad \ell_{h_{T\!-\!1}} \one \leq \sum_{\asm{0.6}{$\theta$}} R_{h_{T\!-\!1}}^\theta  \zbf_{h_{T\!-\!1}}^{\theta} \!\!+\! \sum_{\asm{0.6}{$h_T$}}\! \paren{\ell_{h_T} \,\ebf_{u_{T\!-\!1}}}},
    \label{constr: two stage defender l_h_T-1 optimality}
    \\
    & \as{\forall \, h_T \!\in \!\mathcal{H}_T,}\nonumber \\
    & \as{\quad \one^\top \zbf_{h_T}^\theta\!\! = \!\asm{1}{$\displaystyle\P_{\!\!\!s_{T\!-\!1}\!, u_{T\!-\!1}}^{s_T}$} z_{h_{T\!-\!1}}^\theta\!(v_{T\!-\!1})}, \as{\quad \zbf_{h_T}^\theta \!\geq\! \zero, \quad\quad \forall \, \theta \!\in \!\Theta,}
    \label{constr: two stage defender z_h_T probability}\\
    & \as{\quad \ell_{h_T} \, \one \leq  \sum_{\asm{0.6}{$\theta$}} R_{h_T}^\theta  \zbf_{h_T}^\theta.}
    \label{constr: two stage defender l_h_T optimality}
    \\[-20pt]\nonumber
\end{alignsize}
\end{subequations}
The scalar variable $\ell_{h_T}$ in~\eqref{constr: two stage defender l_h_T optimality} captures the performance at stage $T$.
Note that $\ell_{h_T}$ is scaled by the probability of realizing the history $h_T\!=\!(h_{T-1}, u_{T-1}, v_{T-1}, s_T)$, since this probability is encoded in $\zbf_{h_T}^\theta$ as in~\eqref{constr: two stage defender z_h_T probability}. 
The stage-($T\!-\!1$) value at $\ell_{h_{T-1}}$ is linked to stage-$T$ values through the dynamic-program-like constraint~\eqref{constr: two stage defender l_h_T-1 optimality}.
On the right-hand side of~\eqref{constr: two stage defender l_h_T-1 optimality}, the first summation captures the current performance, while the second represents the expected future performance.
Since the transition probabilities are already incorporated in $\ell_{h_T}$, the expression sums directly over all possible next histories.
%
%
Moreover, constraint~\eqref{constr: two stage defender l_h_T-1 optimality} is an element-wise vector constraint, where each row corresponds to a possible \attacker{} action $u_{T-1}$. 
Combined with the maximization objective~\eqref{obj: two stage Defender}, \eqref{constr: two stage defender l_h_T-1 optimality} ensures optimal \defender{} strategy against all \attacker{} actions for this two-stage problem.




\subsection{General T-stage LP}

With the $\zbf$-variables being defined analogously to \eqref{eqn:two-stage-defender-change-of-var} as
\begin{equation*}
    \zbf_{h_{0}}^\theta = b_0(\theta) \, \taubf_{h_0}^\theta, \quad
    \zbf_{h_t}^\theta = \P_{\!\!\!s_{t-1}, u_{t-1}}^{s_t} \, z_{h_{t-1}}^\theta\!(v_{t-1}) \, \taubf_{h_{t}}^\theta,
\end{equation*}
we can generalize the two-stage LP in~\eqref{LP:two-stage-Defender-LP} to the following generic $T$-stage \defender{} LP for $V^*\!\paren{h_t, \bbf_0}$ using backward induction
\begin{empheqboxed}
\begin{subequations}
\label{LP:generic-defender-LP}
    \begin{align}
    V^*\!\paren{h_0, \bbf_0}=&\!\!\!\!\! \max_{\asm{0.8}{$\substack{ \zbf_{h_t}\!, \,\ell_{h_t} \\ \forall \, t \in \bracket{0:T}, \,h_t \in \mathcal{H}_t }$}}\!\!\! \ell_{h_0}
    \label{obj:generic-defender-LP}
    \\
    \mathrm{s.t.} \quad &  \forall \, \theta \in \Theta, \nonumber \\
    & \quad  \one^\top \zbf_{h_0}^\theta = b_0(\theta), \quad \zbf_{h_0}^\theta \geq \zero,
    \label{constr:defender-lp-prop-0}
    \\
    & \quad  \ell_{h_0} \one\leq \sum_\theta R_{h_0}^\theta \,\zbf_{h_0}^\theta + \sum_{h_1} \paren{\ell_{h_1}\, \ebf_{u_0}}, \label{constr:defender-lp-opt-0}\\[+5pt]
    & \forall \, t \in \bracket{1\!:\!T\!-\!1}, \, \forall \, h_t \in \mathcal{H}_t, \, \forall \, \theta \in \Theta,\nonumber \\
    &  \quad \one^\top \zbf_{h_t}^\theta = \scalebox{1}{$\P_{\!\!\!s_{t\!-\!1}, u_{t\!-\!1}}^{s_t} z_{h_{t\!-\!1}}^\theta\!(v_{t\!-\!1})$}, ~ \zbf_{h_t}^\theta \geq \zero,
    \label{constr:defender-lp-prop-t}
    \\
    & \quad  \ell_{h_t} \one \leq \sum_\theta R_{h_t}^\theta\, \zbf_{h_t}^\theta + \scalebox{1}{$\displaystyle \sum_{h_{t+1}} \paren{\ell_{h_{t+1}}\, \ebf_{u_t}}$}, \label{constr:defender-lp-opt-t}\\[+5pt]
    & \forall \, h_T \in \mathcal{H}_T, \, \forall \, \theta \in \Theta,\nonumber \\
    & \quad  \one^\top \zbf_{h_T}^\theta = 
    \scalebox{1}{$\displaystyle \P_{\!\!\!s_{T\!-\!1}, u_{T\!-\!1}}^{s_T} z_{h_{T\!-\!1}}^\theta \!\!\paren{v_{T\!-\!1}}$}
    , ~ \zbf_{h_T}^\theta \geq \zero,
    \label{constr:defender-lp-prop-T}\\
    & \quad \ell_{h_T}\, \one\leq \sum_{\theta} R_{h_T}^\theta \,\zbf_{h_T}^\theta, \label{constr:defender-lp-opt-T}
\end{align}
\end{subequations}
\end{empheqboxed}


The \defender{} policy is reconstructed from the optimal solution $\zbf^*$ of \eqref{LP:generic-defender-LP} via the following normalization
\begin{equation*}
\taubf_{h_t}^{\theta*} =
\begin{cases}
\displaystyle{\frac{\zbf_{h_t}^{\theta*}}{\one^\top \, \zbf_{h_t}^{\theta*}}}, & \text{if } \one^\top \zbf_{h_t}^{\theta*}\neq 0,\\
\displaystyle{\frac{1}{|\V|}}, & \text{otherwise}.
\end{cases}
\end{equation*}

\subsection{Interpretation of the Defender's LP}

The LP formulation~\eqref{LP:generic-defender-LP} consists of two types of constraints. 
Constraints \eqref{constr:defender-lp-opt-0}, \eqref{constr:defender-lp-opt-t}, and \eqref{constr:defender-lp-opt-T} encode optimality. 
%
%
Constraints \eqref{constr:defender-lp-prop-0}, \eqref{constr:defender-lp-prop-t}, and \eqref{constr:defender-lp-prop-T} propagate the joint distribution over types and history realizations. 
%

The LP formulation \eqref{LP:generic-defender-LP} offers insight into the asymmetric information structure in DRAG problems. 
The $\zbf$ variables encode joint distributions over types and histories, showing that the informed \defender{} plans over distributions of game trajectories.
By shaping these trajectories, the \defender{} influences the uninformed \attacker{}’s belief updates and decisions.

\subsection{Computation Complexity}

We next characterize the size of the LP in~\eqref{LP:generic-defender-LP}.
For each history $h_t$ at time $t \in \bracket{0:T}$, the LP contains one scalar variable $\ell_{h_t}$ and $|\Theta|$ vectors $\zbf_{h_t}^\theta \in \mathbb{R}^{|\V|}$.
Since $|\V|=|\Theta|$, the total number of scalar variables is
\begin{alignsize}{0.9}
    \as{N_{\mathrm{var}}=\sum_{t=0}^T \sum_{h_t \in \mathcal{H}_t} \paren{1 + |\Theta|^2}=\sum_{t=0}^T \bracket{|\mathcal{H}_t|\cdot \paren{1+|\Theta|^2}}}
    \label{eqn:generic-defender-LP-num-of-var}
\end{alignsize}
variables.
For each $h_t$, the probability-flow constraints contribute $|\Theta|$ equalities, while the optimality constraint contributes $|\U\!\paren{s_t}\!|$ inequalities.
Thus, excluding the non-negativity bounds on $\zbf$, the number of constraints is
%
\begin{alignsize}{1}
    \as{N_{\mathrm{constr}}=\sum_{t=0}^T \sum_{h_t \in \mathcal{H}_t} \paren{|\Theta|+|\,\U\!\paren{s_t}\!|}.}
    \label{eqn:generic-defender-LP-num-of-constr}
\end{alignsize}
The $\zbf$-variables additionally impose $\sum_{t=0}^T \paren{|\mathcal{H}_t|\cdot|\Theta|^2}$ non-negativity bounds.
%
%

Let $|\U|_{\max} = \displaystyle \max_{s \in \S} |\,\U\!\paren{s}\!|$,
then, each stage has at most $|\U|_{\max} |\Theta| |\S|$ history branches.
Hence $|\mathcal{H}_t|$ is upper bounded by $ |\U|_{\max} |\Theta| |\S|$.
%
The numbers of variables and constraints grow exponentially with the horizon $T$ in the worst case, but polynomially with $|\U|_{\max}$, $|\Theta|$, and $|\S|$ for fixed $T$.
%
%
%

\section{Attacker's Linear Program}

By strong duality, the \attacker{}'s equilibrium strategy $\sigma^*$ can be solved from the dual of the \defender{} LP in~\eqref{LP:generic-defender-LP}, following a max-min formulation of $V^*(h_t, \bbf_t)$.
Similar to the \defender{} LP, we use the change of variables $\etabf_{h_0} \!=\! \sigmabf_{h_0}$ and $\etabf_{h_t} \!\!=\! \eta_{h_{t-1}}\!\!\paren{u_{t-1}} \sigmabf_{h_t}$, yielding the following $T$-stage \attacker{} LP 


%
%


\begin{empheqboxed}
\begin{subequations}
\label{LP:generic-attacker-LP}
\begin{alignsize}{1}
    \as{V^*\!\paren{h_0, \bbf_0}=}&\as{\!\!\!\!\min_{\asm{0.8}{$\substack{\etabf_{h_t}, \, \qbf_{h_t}\\ \forall \, t \in \bracket{0:T},\, h_t\in \mathcal{H}_t}$}} \sum_{\theta} b_0\!\paren{\theta} q_{h_0}\!\paren{\theta}} \label{eqn:attacker-lp-objective}\\
    \as{\mathrm{s.t.~~}} & \as{\forall \, \theta\! \in\! \Theta, \, \forall \, v_0 \!\in\! \V,} \nonumber \\
    & \as{\quad \one^\top \etabf_{h_0} = 1, \quad \etabf_{h_0} \geq \zero,} 
    \label{eqn:attacker-lp-prop-0}
    \\
    & \as{\quad q_{h_0}\!\paren{\theta} \geq \ebf_{v_0}^\top R_{h_0}^{\theta^\top} \etabf_{h_0} + \!\!\sum_{s_1, u_0}\!\paren{\P_{\!\!\!s_0, u_0}^{s_1} q_{h_1}(\theta)},} 
    \label{eqn:attacker-lp-opt-0}
    \\[+5pt]
    & \as{\forall \, t \!\in \!\bracket{1\!:\!T\!-\!1}, \, \forall \, h_t \!\in\! \mathcal{H}_t,\, \forall \, \theta\! \in\! \Theta, \, \forall \, v_t \!\in\! \V,} \nonumber \\
    & \as{\quad \one^\top \etabf_{h_t} = \eta_{h_{t-1}}\!\!\paren{u_{t-1}}, \quad \etabf_{h_t} \geq \zero,}
    \label{eqn:attacker-lp-prop-t}
    \\
    &\as{\quad q_{h_t}\!\paren{\theta} \geq \ebf_{v_t}^\top R_{h_t}^{\theta^\top} \etabf_{h_t} + \!\!\!\!\sum_{s_{t+1}, u_t} \!\!\!\!\paren{\P_{\!\!\!s_t, u_t}^{s_{t+1}} q_{h_{t+1}}\!\paren{\theta}},}
    \label{eqn:attacker-lp-opt-t}
    \\[+5pt]
    & \as{\forall \, h_T \!\in\! \mathcal{H}_T, \, \forall \, \theta \!\in\! \Theta, \, \forall \, v_T\! \in\! \V,}\nonumber \\
    & \as{\quad \one^\top \etabf_{h_T} = \eta_{h_{T\!-\!1}}\!\!\paren{u_{T\!-\!1}}, \quad \etabf_{h_T} \geq \zero,}
    \label{eqn:attacker-lp-prop-T}\\
    & \as{\quad q_{h_T}(\theta) \geq  \ebf_{v_T}^\top \, R_{h_T}^{\theta^\top} \, \etabf_{h_T}.}
    \label{eqn:attacker-lp-opt-T}
\end{alignsize}
\end{subequations}
\end{empheqboxed}



%
Both \eqref{LP:generic-defender-LP} and \eqref{LP:generic-attacker-LP} compute $V^*\!\paren{h_0, \bbf_0}$, but in two different ways.
The \defender{}'s and \attacker{}'s equilibrium strategies are recovered by normalizing the optimal solutions of \eqref{LP:generic-defender-LP} and \eqref{LP:generic-attacker-LP}, respectively.
Since \eqref{LP:generic-attacker-LP} is the dual of \eqref{LP:generic-defender-LP}, it has $N_{\mathrm{constr}}$ variables and $N_{\mathrm{var}}$ constraints, excluding variable-sign restrictions.
%
Unlike the $\zbf$ variables, $\etabf$ encodes only the \attacker{}’s action distribution along each game trajectory,
%
%
reflecting the \attacker{}’s informational disadvantage. 
Since belief updates~\eqref{eqn:belief-update-rule} are determined solely by the \defender{}’s strategy, the \attacker{} neither controls nor plans how its belief evolves, but instead optimizes performance aggregated across 
all game types.
%
Due to strong duality, the LPs in~\eqref{LP:generic-defender-LP} and~\eqref{LP:generic-attacker-LP} constructed from the dynamic programs ensure the existence of PBNE in DRAG, as specified in Theorem~\ref{theorem:PBNE-existence}.
%
\begin{theorem}[Existence of PBNE]
\label{theorem:PBNE-existence}
The DRAG can be solved as a pair of primal-dual linear programs, which admits a unique and finite game value $V^*\!\paren{h_0, \bbf_0}$ due to strong duality.
Through normalization, the LPs' optimal solutions constitute a PBNE.
\end{theorem}

\section{Numerical Experiments}

\begin{figure}[b!]
    \vspace{-0.2in}
    \centering
    \includegraphics[width=0.62\linewidth]{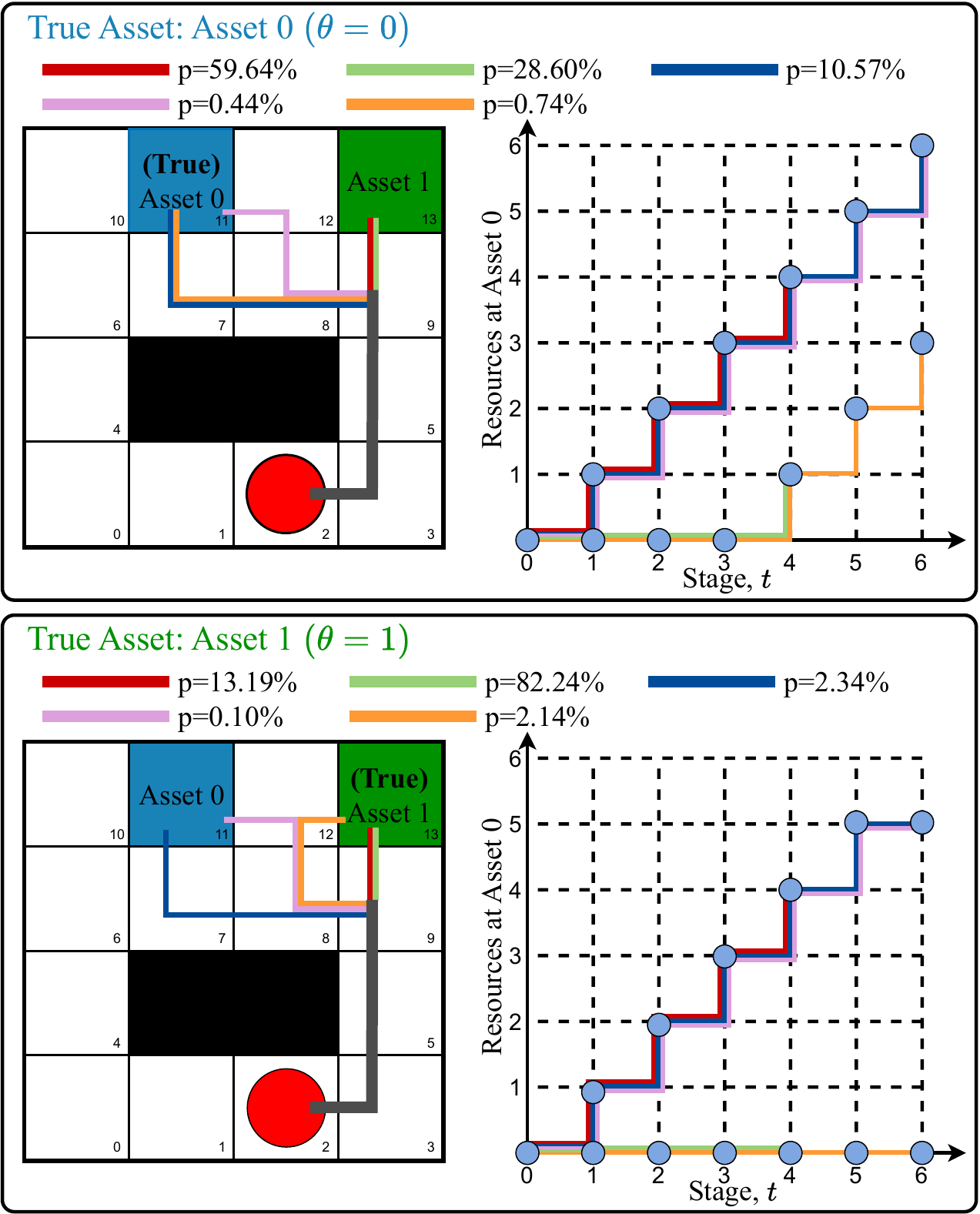}
    \caption{
    Histories of a $4\!\times\!4$ DRAG.
    The label $p$ denotes the probability of the game history $h_t$ depicted in the corresponding color.
        \emph{Top} \& \emph{Bottom:} \attacker{} trajectories and \defender{} allocation when Asset~0~\&~1 is selected.
    }
    \label{fig: DRAG True Goal 0 Trajectory}
\end{figure}
We evaluated the proposed LP approach on the $4\!\times\!4$ grid DRAG in Fig.~\ref{fig: DRAG True Goal 0 Trajectory}, where the two candidate assets are marked in green and blue, and the prior is set to $\bbf_0\!=\![0.2,\,0.8]$.
%
%
The running reward for a type-$\theta$ Defender is defined as
\vspace{-0.03in}
\begin{equation*}
r(h_t,\theta,u_t,v_t)=\mathds{1}(v_t=\theta)\,\mathds{1}(\Theta\cap s_{0:t}=\emptyset),
\vspace{-0.02in}
\end{equation*}
i.e., the \defender{} receives reward only when allocating to the true asset before the \attacker{} reaches any candidate asset.
%

The game terminates when the \attacker{} either (i) fails to reach any asset by the horizon $T=10$, or (ii) reaches one asset at the terminal time $\hat{t} \leq T$.
%
At time $\hat{t}$, we assign the additional terminal reward
\vspace{-0.05in}
\[
    r_{\hat{t}} = -m\,\mathds{1}(s_{\hat t} = \theta).
    \vspace{-0.05in}
\]
That is, if the Attacker reaches the true asset, the Defender is penalized with the Attacker threat level $m=25$. 
Figure~\ref{fig: DRAG True Goal 0 Trajectory} presents the possible game histories $h_t$ under the strategy profile $\paren{\sigma^*, \tau^*}$ obtained from the LPs in~\eqref{LP:generic-defender-LP} and \eqref{LP:generic-attacker-LP}.
Since $h_t$ contains both the \attacker{} trajectory and the \defender{} allocation history, we visualize them separately for clarity. 
With only two assets, the \defender{}'s allocation is characterized by resources at Asset~0. 
For an arbitrary $h_t$, the corresponding \attacker{} trajectory and the \defender{} allocation are plotted using the same color.

\subsection{Deceptive Behaviors}
\begin{figure*}[b]
    \centering
    \includegraphics[width=\linewidth]{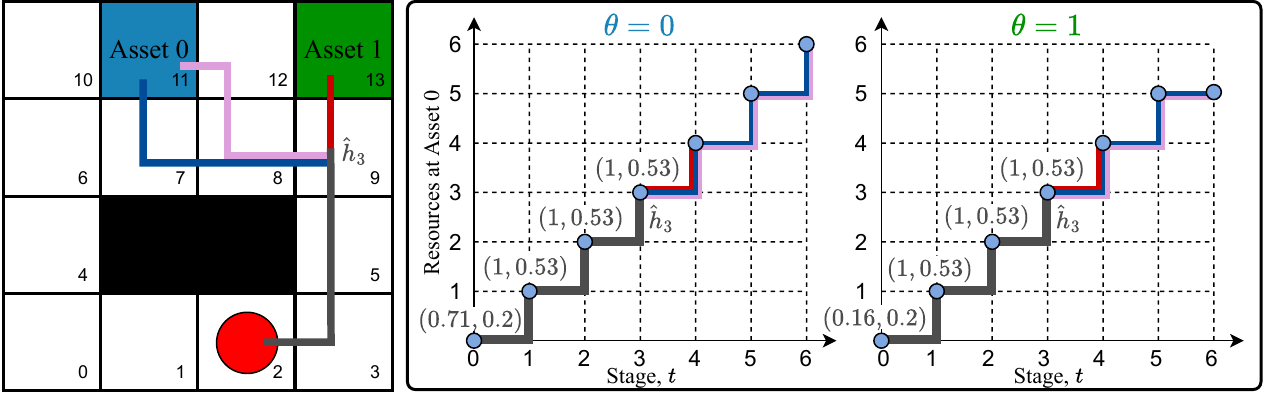}
    \vspace{-0.2in}
    \caption{Illustration of deception.
    \emph{Left:} Attacker trajectories;
    \emph{Middle}
    \& \emph{Right:} Type-0~\&~1 Defender allocation histories.}
    \label{fig: highlighted gameplay trajectories}
\end{figure*}

We first illustrate how $\paren{\sigma^*, \tau^*}$ induces belief manipulation and leads to emergent deceptive behavior.
In Fig.~\ref{fig: highlighted gameplay trajectories}, we consider all trajectories following the parent history $\hat{h}_3$, under which the \attacker{} follows the state sequence $(2,3,5,9)$ and the \defender{} allocates to Asset~0 throughout. 
Along $\hat{h}_3$, the evolution of the \defender{}'s policy and the \attacker{}'s belief is characterized by $\left\{(\tau_{h_t}^\theta(0), b_t(0))\right\}_{t=0}^3$ in Fig.~\ref{fig: highlighted gameplay trajectories}.
At time $t\!=\!0$, the \defender{} randomizes its allocation with policy $\taubf_{h_0}^0\!\!=\!\![0.71, \,0.29]$, $\taubf_{h_0}^1 \!\!=\!\! [0.16,\, 0.84]$,
This randomized strategy induces the posterior \attacker{} belief $\bbf_1 \!\!=\!\! [0.53,\, 0.47]$, which is later preserved by the type-invariant policy $\taubf_{h_t}^\theta \! = \![1,0]$ until reaching state~$9$, strategically positioning the \attacker{} at an indifference point.
At state $9$, the Attacker faces two possible actions: moving \texttt{Left} toward Asset~0 or \texttt{Up} toward Asset~1, each yielding different payoffs depending on the type $\theta$. 
Specifically, moving to Asset~0 yields payoffs $(-19,\,1)$ under types $(0,1)$, while moving to Asset~1 yields $(4,\,-25)$.
Under the induced belief $[0.53,\,0.47]$, the expected payoff of moving to either asset is identical, i.e.
\begin{alignsize*}{0.95}
\as{0.53 \times (-19) + 0.47 \times 1
=
0.53 \times 4 + 0.47 \times (-25)}.
\end{alignsize*}
Therefore, the Attacker is indifferent among all actions at state $9$, which is the consequence of the \defender{}’s deliberate randomization, which strategically shapes the \attacker{}’s ambiguous belief $[0.53, 0.47]$.
However, this randomization may appear ``suboptimal'', since the type-0 \defender{} allocates to the decoy Asset 1 with a probability of 0.29, effectively ``wasting" resources.
As we will show later, inducing such ambiguity at the critical \attacker{} decision point in fact maximizes \defender{} performance, thus constituting \emph{purposeful deception}.
We use the term \emph{purposeful deception} to describe deceptive behavior that emerges endogenously from the PBNE when influencing the \attacker{}'s belief and subsequent actions improves the \defender{}'s equilibrium performance.
In contrast, optimizing a separate deception heuristic may encourage unnecessary deceptive behavior that degrades overall performance, resulting in ``deception for the sake of deception''.
%
%

\subsection{Effectiveness of Deception}

The LP-generated strategy profile $\paren{\sigma^*, \tau^*}$ in Fig.~\ref{fig: DRAG True Goal 0 Trajectory} leads to an expected performance
\(
V_{\mathrm{LP}}^*(s_0,\bbf_0)\!=\!-16.68
\).
To quantify the Defender's informational advantage, we compared with the full-information game outcome $V_{\mathrm{FI}}^*(s_0, \bbf_0)$,
where Nature selects the true asset via $\bbf_0$ and publicly announces it.
In the type-$\theta$ full-information game, the optimal Attacker strategy is the shortest path to the true asset $\theta$, while the Defender best responds by allocating to $\theta$ at every step. 
Such a strategy pair leads to the full-information value
\begin{alignsize*}{0.95}
\as{V_{\mathrm{FI}}^*(s_0, \bbf_0)
=
0.2 \times (-19)+0.8 \times (-21)
=
-20.6}.
\end{alignsize*}
With the \emph{value of information} (VoI)~\cite{rostobaya2025deceptive} defined in~\eqref{eqn:VoI}, we can capture the Defender’s performance gain due to informational advantage:
\vspace{-0.05in}
\begin{alignsize}{.95}
\as{\mathrm{VoI}\paren{s_0, \bbf_0} = \frac{V_{\mathrm{LP}}^*\paren{s_0, \bbf_0} - V_{\mathrm{FI}}^*\paren{s_0, \bbf_0}}{|V_{\mathrm{FI}}^*\paren{s_0, \bbf_0}|}}.    
\label{eqn:VoI}
\end{alignsize}

For the $4\times4$ DRAG, the performance gap between the two values is $3.92$, corresponding to a VoI of $19\%$.
This indicates that, by purposefully allocating to the decoys, the \defender{} manipulates the Attacker’s belief, leading to a $19\%$ improvement with respect to the full-information setting.

%

\subsection{Equilibrium Validation}

We next verify that $(\sigma^*,\tau^*)$ constitutes a PBNE by showing that neither player benefits from unilateral deviation. 
%
%

\begin{figure*}
\vspace{-25pt}
    \centering
    \captionof{table}{Performances under unilateral \attacker{} deviation.}
    \label{tab:value-under-attacker-deviation}
    \begin{minipage}{\linewidth}
    \vspace{+5pt}
    \centering
    \begin{tabular}{l l l l l}
     \toprule
    $V^*\!\paren{s_0, \bbf_0, \sigma, \tau^*}$
    \quad\quad
    & \texttt{LP-A}, $\sigma^*$ \quad\quad & \texttt{RS-A} \quad\quad & \texttt{HPSP-A}\quad\quad & \texttt{LPSP-A}\\
    \midrule
    \texttt{LP-D}, $\tau^*$ \quad\quad& \textbf{-16.68} & -8.32 & \textbf{-16.68} & -0.85\\
    \midrule
    \end{tabular}
    \end{minipage}
\vspace{-0.4in}
\end{figure*}
\paragraph{Attacker Deviation}
We consider the following alternative \attacker{} strategies:
\texttt{RS-A} samples actions uniformly;
\texttt{HPSP-A} and \texttt{LPSP-A} follow a shortest path to the highest-prior and the lowest-prior assets, respectively.
%
Table~\ref{tab:value-under-attacker-deviation} reports the performances of the alternative Attacker strategies against the Defender strategy $\tau^*$. 
As expected, all deviations yield values no smaller than $V_{\mathrm{LP}}^*(s_0,\bbf_0)$, confirming that the (minimizing) Attacker cannot improve by deviating from $\sigma^*$.
Notably, \texttt{HPSP-A} attains the same performance as $\sigma^*$.
However, the \defender{} can best respond against \texttt{HPSP-A} by consistently allocating to the true asset, thus exploiting \texttt{HPSP-A}.
This yields a value of $-16$, strictly improving the Defender’s payoff relative to the equilibrium. 
%
\vspace{-25pt}
\begin{figure*}[htb]
\centering
\captionof{table}{
Performances under unilateral \defender{} deviation.
}
\label{tab:value-under-defender-deviation}
\begin{minipage}{\linewidth}
\vspace{+5pt}
    \centering
    \begin{tabular}{l l l l l l l}
    \midrule
    $V^*\!\paren{s_0, \bbf_0, \sigma^*, \tau}$\quad\quad\quad
    & \texttt{LP-D}, $\tau^*$\quad\quad\quad & \texttt{RS-D} \quad\quad\quad & \texttt{TC-D}\quad\quad\quad & \texttt{TO-D} \quad\quad\quad & \texttt{C0-D}\quad\quad\quad & \texttt{C1-D}\\
    \midrule
    \texttt{LP-A}, $\sigma^*$ & \textbf{-16.68} & -16.85 & \textbf{-16.68} & -16.88 & -16.80 & -16.76\\
    \bottomrule
\end{tabular}
\end{minipage}
\vspace{-0.3in}
\end{figure*}

\paragraph{Defender Deviation}

We experimented with the Defender alternatives: \texttt{RS-D} uniformly samples actions; \texttt{TC-D} and \texttt{TO-D} respectively allocate to true and decoy assets.
; \texttt{C0-D} and \texttt{C1-D} respectively allocate to Asset~0 and Asset~1 regardless of the selected true asset.
%
Table~\ref{tab:value-under-defender-deviation} shows that all deviations yield performances no greater than $V_{\mathrm{LP}}^*(s_0,\bbf_0)$, validating the optimality of $\tau^*$.
Although \texttt{TC-D} achieves the same performance as $\tau^*$, the \attacker{} can exploit it by moving toward the true asset identified immediately after the first observation, yielding a value of $-20.6$ that is strictly lower
than $V_{\rm{LP}}^*\!\paren{s_0, \bbf_0}$.

\section{Conclusion}

We introduced the Deceptive Resource Allocation Game (DRAG) and studied purposeful deception under a Bayesian game framework.
While the informed player does not fully control the observed signal, we showed that DRAG still admits a linear program for computing Perfect Bayesian Nash Equilibria.
Numerical results show that purposeful deception arises by shaping beliefs to influence decisions at critical states, rather than simply maximizing inference bias or uncertainty.
Future work includes extension to two-sided information asymmetry and larger-scale environments.

\noindent\textbf{Acknowledgements:}
This work was supported by the U.S. Army Research Laboratory (ARL) through the Distributed and Collaborative Intelligent Systems and Technology (DCIST) Collaborative Research Alliance (CRA) under Grant W$911$NF-$17$-$2$-$0181$ and Office of Naval Research (ONR) award N$00014$-$23$-$1$-$2308$.

\noindent\textbf{Disclosure of Interests:}
The authors have no competing interests to declare that are relevant to the content of this article.

\newpage

\begin{appendix}
\section{Appendix}
\subsection{Proof of Proposition~\ref{prop:belief-update-satisfies-belief-definition}}
\label{appendix:belief-update-justification}
\begin{proof}
    For $t \!\in \!\bracket{1:T}$, given the observed history $h_t \!= \!\paren{h_{t-1}, u_{t-1}, v_{t-1}, s_t}$, we will show \eqref{eqn:belief-update-rule} outputs $b_t\!\paren{\theta}$ under Assumption~\ref{assump:belief-cond-indep}.

    By Definition~\ref{def:belief-definition}, the numerator of \eqref{eqn:belief-update-rule} can be interpreted as
    \vspace{-5pt}
    \begin{alignsize*}{0.9}
        \as{b_{t-1} \!\paren{\theta} \tau_{h_{t-1}}^{\theta}\!\paren{v_{t-1}}} &  ~\as{= \P^\tau \!\paren{\theta \mid h_{t-1}} \P^\tau\!\paren{v_{t-1} \mid h_{t-1}, \theta}}\\
        & \as{= \frac{\P\!\paren{h_{t-1}, \theta}}{\P\!\paren{h_{t-1}}} \frac{\P\!\paren{v_{t-1}, h_{t-1}, \theta}}{\P\!\paren{h_{t-1}, \theta}}}\\
        & \as{= \P^\tau\!\paren{v_{t-1}, \theta \mid h_{t-1}}.}
        \\[-20pt]
    \end{alignsize*}
    The denominator of \eqref{eqn:belief-update-rule} is a marginalization of the numerator, which can be interpreted as
    \vspace{-3pt}
    \begin{equation*}
        \asm{0.9}{$\displaystyle \sum_{\theta'} b_{t-1}\paren{\theta'} \tau_{h_{t-1}}^{\theta'}\paren{v_{t-1}} = \P^\tau\paren{v_{t-1} \mid h_{t-1}}.$}\vspace{-5pt}
    \end{equation*}
    Therefore, \eqref{eqn:belief-update-rule} can be expressed as follows
    \vspace{-5pt}
    \begin{alignsize*}{0.9}
        \as{\frac{b_{t\!-\!1}\!\paren{\theta} \tau_{h_{t\!-\!1}}^\theta\!\!\paren{v_{t\!-\!1}}}{\sum_{\theta^{'}} b_{t\!-\!1}\!\paren{\theta^{'}} \tau_{h_{t\!-\!1}}^{\theta^{'}}\!\!\!
        \paren{v_{t\!-\!1}}}}~ & \as{= \frac{\P^{\tau}\!\paren{v_{t\!-\!1}, \theta \mid h_{t\!-\!1}}}{\P^{\tau}\!\paren{v_{t\!-\!1} \mid h_{t\!-\!1}}}}\\[-3pt]
        &\as{ = \frac{\P\!\paren{v_{t\!-\!1}, \theta, h_{t\!-\!1}}}{\P\!\paren{h_{t\!-\!1}}} \frac{\P\!\paren{h_{t\!-\!1}}}{\P\!\paren{v_{t\!-\!1}, h_{t\!-\!1}}}}\\[-3pt]
        & \as{= \P^{\tau}\!\paren{\theta \mid h_{t\!-\!1}, v_{t\!-\!1}}}\\[-3pt]
        &\as{ = b_t\!\paren{\theta},}\\[-20pt]
    \end{alignsize*}
    where the last equality is a result of Assumption~\ref{assump:belief-cond-indep}.
    \hfill $\blacksquare$
\end{proof}

\subsection{Objective Upper-Boundedness}
\label{appendix:lemma-1-proof}
\begin{lemma}
\label{lemma:two-stage-inner-max-obj-upper-bound}
For a fixed history $h_{T-1}$, define
\vspace{-5pt}
\begin{subequations}
    \label{eqn:two-stage-bar-r-def}
        \begin{alignsize}{1}
            \as{\bar{r}_{T-1}} & \as{\coloneqq \!\!\!\max_{\asm{0.6}{$\theta,\! u_{T\!-\!1},\!v_{T\!-\!1}$}}\!r\!\paren{h_{T\!-\!1}, \theta, u_{T\!-\!1}, v_{T\!-\!1}},}
            \label{eqn:bar_r_T-1_def}
            \\
            \as{\bar{r}_T} & \as{\coloneqq  \max_{\asm{0.6}{$h_T,\! \theta,\! u_{T},\! v_T$}} r\!\paren{h_T, \theta, u_T, v_T}.}
            \label{eqn:bar_r_T_def}\\[-20pt]\nonumber
        \end{alignsize}
    \end{subequations}
    Then, every feasible solution of~\eqref{LP:two-stage-inner-max} satisfies 
    \begin{alignsize}{1}
        \as{\sigmabf_{h_{T\!-\!1}}^\top\bigg\{\sum_{\asm{0.6}{$\theta$}} R_{h_{T\!-\!1}}^\theta \zbf_{h_{T\!-\!1}}^\theta + \bigg[\sum_{\asm{0.6}{$v_{T\!-\!1},\! s_T$}} \!\ell_{h_T}\bigg]_{\asm{0.6}{$u_{T\!-\!1}$}}\bigg\}} & \as{\leq \bar{r}_{T\!-\!1} + \bar{r}_{T}.}\label{eqn:two-stage-inner-max-obj-upper-bound}
        \\[-20pt] \nonumber
    \end{alignsize}
\end{lemma}
\begin{proof}
Since the state and action spaces and the horizon are finite, the game tree of DRAG is finite.
Hence, the maxima $\bar{r}_{T-1}$ and $\bar{r}_T$ in~\eqref{eqn:two-stage-bar-r-def} are finite and attainable.

Under a fixed $\sigmabf_{h_{T-1}}$, we will show the objective~\eqref{obj:two-stage-inner-max} is bounded above by $\bar{r}_{T-1} + \bar{r}_T$ for all feasible solutions.
%
From the first part of~\eqref{obj:two-stage-inner-max}, we have
\begin{alignsize}{0.9}
    \as{\sigmabf_{h_{T-1}}^\top \sum_{\as{\theta}} R_{h_{T-1}}^\theta \zbf_{h_{T-1}}^\theta} & \as{= \!\!\!\!\sum_{\asm{0.8}{$\theta, u_{T-1}, v_{T-1}$}} \!\!\!\!\!\!\sigma_{h_{T-1}}\!\paren{u_{T-1}} r\!\paren{h_{T-1}, \theta, u_{T-1}, v_{T-1}} z_{h_{T-1}}^\theta\!\paren{v_{T-1}}} \nonumber\\[-3pt]
    & \as{\leq \bar{r}_{T-1} \sum_{\asm{0.8}{$u_{T-1}$}}\Big( \sigma_{h_{T-1}}\!\paren{u_{T-1}} \sum_{\asm{0.8}{$\theta$}} \one^\top \zbf_{h_{T-1}}^\theta}\Big) \label{eqn:lemma-1-eqn-1} \\[-3pt]
    & \as{= \bar{r}_{T-1} \sum_{\asm{0.8}{$u_{T-1}$}}\Big( \sigma_{h_{T-1}}\!\paren{u_{T-1}} \sum_{\asm{0.8}{$\theta$}} b_{T-1}\!\paren{\theta}}\Big)\label{eqn:lemma-1-eqn-2}\\[-5pt]
    &\as{= \bar{r}_{T-1},} \nonumber
\end{alignsize}
where~\eqref{eqn:lemma-1-eqn-1} follows from the definition of $\bar{r}_{T-1}$ in~\eqref{eqn:bar_r_T-1_def} and $\zbf_{h_{T\!-\!1}}^\theta\!\!\geq \!\zero$;
the last two equalities use \eqref{constr:two-stage-inner-max-z_h_T-1-probability}, $\sum_\theta b_{T\!-\!1}\!\paren{\theta}\!\!=\!1$, and $\one^\top \sigmabf_{h_{T\!-\!1}}\!\!=\!1$.

For any successor history $h_T\!=\!\paren{h_{T-1}, u_{T-1}, v_{T-1}, s_T}$, notice constraint~\eqref{constr:two-stage-inner-max-l_h_T-optimality-z} implies, for every $u_T \in \U\!\paren{s_T}$,
\vspace{-10pt}
\begin{alignsize}{0.9}
    & \as{\ell_{h_T} } \as{\leq \ebf_{u_{T}}^\top \sum_{\asm{0.8}{$\theta$}} R_{h_T}^\theta \zbf_{h_T}^\theta} 
    \as{\leq \bar{r}_{T} \sum_{\asm{0.8}{$\theta$}} \one^\top \zbf_{h_T}^\theta}
    \as{=\bar{r}_T \sum_{\asm{0.8}{$\theta$}} \P_{\!\!\!s_{T\!-\!1}, u_{T\!-\!1}}^{s_T} z_{h_{T\!-\!1}}^\theta\!\paren{v_{T\!-\!1}}.}
    \label{eqn:l_h_T_ineq}
    \\[-22pt] & \nonumber 
\end{alignsize}
%
Using~\eqref{eqn:l_h_T_ineq}, the definition of $\bar{r}_T$ in~\eqref{eqn:bar_r_T_def}, and the probability-flow constraints on $\zbf_{h_T}^\theta$ in~\eqref{constr:two-stage-inner-max-z_h_T-1-probability} and~\eqref{constr:two-stage-inner-max-z_h_T-probability}, the second part of~\eqref{obj:two-stage-inner-max} can be expressed as
\vspace{-5pt}
\begin{alignsize}{0.9}
    \as{\sigmabf_{h_{T\!-\!1}}^\top \bigg[\sum_{\asm{0.8}{$v_{T\!-\!1}, s_T$}} \!\!\ell_{h_T}\bigg]_{\asm{0.8}{$u_{T\!-\!1}$}}} &\!\! = \as{\!\!\sum_{\asm{0.8}{$u_{T\!-\!1}, v_{T\!-\!1}, s_T$}}\!\!\!\!\!\!\!\!\!\sigma_{h_{T-1}}\!\paren{u_{T\!-\!1}}\ell_{h_T}}\nonumber\\[-3pt]
    & \as{\leq \bar{r}_{T} \sum_{\asm{0.8}{$u_{T-1}$}}\Big(\sigma_{h_{T\!-\!1}}\!\paren{u_{T-1}}\!\!\sum_{\asm{0.8}{$v_{T\!-\!1}, s_T, \theta$}}\!\!\!\!\!\!\P_{\!\!\!s_{T\!-\!1}, u_{T\!-\!1}}^{s_T} z_{h_{T\!-\!1}}^\theta\!\paren{v_{T\!-\!1}}\Big)}
    \label{eqn:lemma-1-eqn-3}
    \nonumber
    \\[-3pt]
    & \as{= \bar{r}_T \sum_{\asm{0.8}{$u_{T-1}$}}\Big(\sigma_{h_{T-1}}\!\paren{u_{T-1}}\sum_{\asm{0.8}{$s_T$}} \Big(\P_{\!\!\!s_{T\!-\!1}, u_{T\!-\!1}}^{s_T} \sum_{\asm{0.8}{$\theta$}} \one^\top \zbf_{h_{T-1}}^\theta\Big)\Big)}
    \nonumber
    \\[-5pt]
    & \as{= \bar{r}_T.}\nonumber
\end{alignsize}
Combining the inequalities for both parts of~\eqref{obj:two-stage-inner-max} yields the result in~\eqref{eqn:two-stage-inner-max-obj-upper-bound}.
\hfill $\blacksquare$
\end{proof}

\end{appendix}

\medskip
\bibliographystyle{splncs04}
\bibliography{ref.bib}
\end{document}